# Controlling Particulates and Dust in Vacuum Systems

*Lutz Lilje*
DESY Deutsches Elektronen-Synchrotron, Hamburg, Germany

**Abstract**
Particulates and dust can limit accelerator performance in various ways. In this paper the basic properties and mechanisms of contaminations due to particulates are described. With this knowledge countermeasures can be implemented to minimize degradation due to particulates inside the accelerator vacuum systems.

## 1 Dust and particulates

Dust occurs nearly everywhere. Fine parts of matter from various sources are transported in the atmosphere – sometimes over very long distances. While it sounds so mundane, permanent attention is required to limit the effects of dust particulates. In a typical household in Germany 6.2 mg of dust are settling on a square meter per day, see Ref. [1].

Detrimental effects due to dust are known to everybody. Beyond the esthetic effects on furniture surfaces, both humans and technical systems can be heavily affected by dust. Allergic reactions and lung diseases are clearly correlated to certain forms of dust like asthma caused by the feces of dust mites and pneumoconiosis ("Miner's Lung"). A stuck ventilation due to dust has been observed by everybody who has opened a PC after the central processing unit became too hot. Air filters for cars are mandatory to avoid damages to the engines as particulates can significantly enhance wear due to abrasion.

For this paper we will use the words "dust" and "particulates" synonymously. The word "particle" will be used for electrons, protons etc. i.e. the objects a particle accelerator is built for.

The term "particle-free vacuum system" is sometimes used in literature. When people aim for "particle-free" vacuum systems, they mean a vacuum system with the lowest possible count of particulates inside. A truly particulate-free accelerator is difficult – if not impossible – to achieve as several components tend to produce particles during operation.

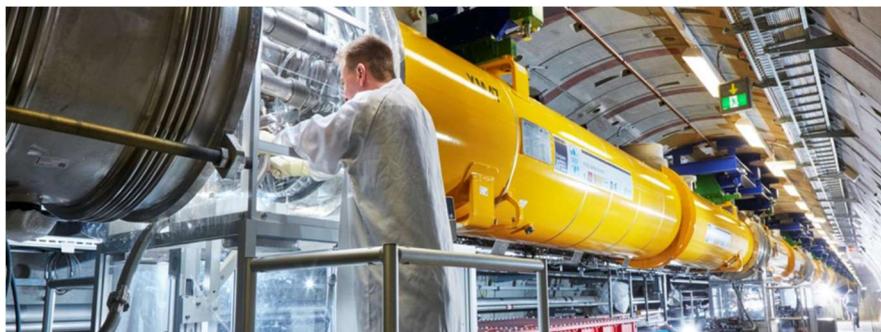

**Fig. 1:** Example for a "particle-free" assembly in the accelerator tunnel of the European XFEL. The interconnection of accelerator modules has to be done in a cleanroom environment to avoid a performance degradation of the superconducting accelerating cavities.

## 2 Problems caused in particle accelerators by dust particulates

In particle accelerators like any other technical installation dust and particulates are not just a nuisance. Several components in accelerators are sensitive to contaminations with particulates. In some cases severe, permanent performance deterioration might occur. Some examples are:
- Components with radiofrequency (RF) or pulsed magnetic fields
  - Superconducting accelerating structures
  - Electron guns
  - Kicker magnets
- Optics components
  - Mirrors
  - Gratings
- Full accelerators

In the following a more detailed discussion on some of these effects will be given.

### 2.1 Superconducting RF accelerating cavities

Superconducting RF cavities are an extremely efficient tool to accelerate particle beams to very high energies. The lower surface resistance resulting in a lower RF power needed for the high accelerating fields as compared to normalconducting accelerating cavities overcompensates the need for cryogenic refrigeration, see Ref. [2]. Superconducting cavities have become a commonplace in particle accelerators. For example, the European XFEL as described in Ref. [3] uses 800 nine-cell 1.3GHz TESLA accelerating structures, which were developed for a linear collider based on superconducting technology, see Ref. [4]. A view into the European XFEL accelerator tunnel is shown in Fig. 1.

#### 2.1.1 Thermal breakdown

Particulate contaminations on the superconductor's surface can quickly increase the overall surface resistance to intolerable levels. As an example in TESLA-style cavities (like for XFEL or ILC) a normalconducting particulate of a about 50 um dissipates enough heat warming up the superconductor at field levels of 25 MV/m to a temperature that a thermal runaway situation occurs. The full cavity becomes normalconducting ("quenches"). In other words: If higher gradients than 25 MV/m are desired, no particulates larger than a few um are allowed on the inner surface.

#### 2.1.2 Field emission

The second limiting effect due to particulate contamination is field emission. The enhanced high electric fields at protrusions from particulates can lead to the emission of electrons from the surface. These unwanted electrons take energy from the accelerating field in the cavity, deposit energy in form of heat on the cavity wall and lead to the emission of radiation. All these effects are undesirable. The energy loss needs to compensated, the additional heating leads to a need of higher cryogenic cooling and the emitted radiation can damage components installed near the accelerating cavities.

#### 2.1.3 Consequences for superconducting cavities

Both effects described before necessitate preparing the surface of an accelerating cavity to level of cleanliness known from semi-conductor industry. The main differences are the significantly larger surface area and the need for an ultra-high vacuum (UHV) for the beam transport.

The inner surface area of accelerating cavities is in the order of square meters. In the case of the European XFEL the inner surface measures about 0.6 m$^2$. It is a challenge to prepare the surface properly to achieve high accelerating gradients. Particulates have to be removed. One efficient tool is to use a jet of high pressure, ultra-pure water. There are several variants of systems for the cleaning of cavities. Descriptions can be found as part of the Proceedings of the Workshops on RF Superconductivity.

As the superconducting cavities are part of the beam line of the accelerator they need to be evacuated to UHV conditions to facilitate beam transport. As a consequence reliable vacuum connections in a cryogenic environment need to be made. Typically, metal flanges and gaskets are used. Screws and nuts are an unavoidable part of this kind of flange systems. When screws are fastened, a production of particulates is unavoidable. Assembly procedures have to take this into account and need to ensure that none of the particulates produced during assembly reaches the inner surface of the accelerating cavity. Of course, this problem also occurs in all other vacuum equipment where metal flanges and gaskets are used.

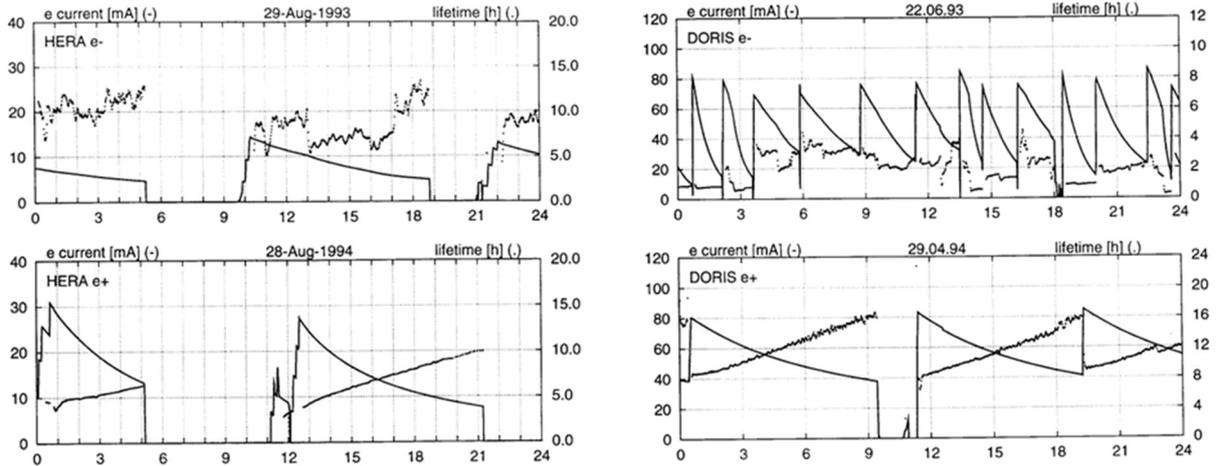

**Fig. 2**: During the operation of HERA (left) and DORIS (right) with electrons (top) sudden changes in the beam lifetime have been observed. Operation of these accelerators with positrons (bottom) has not shown this problem.

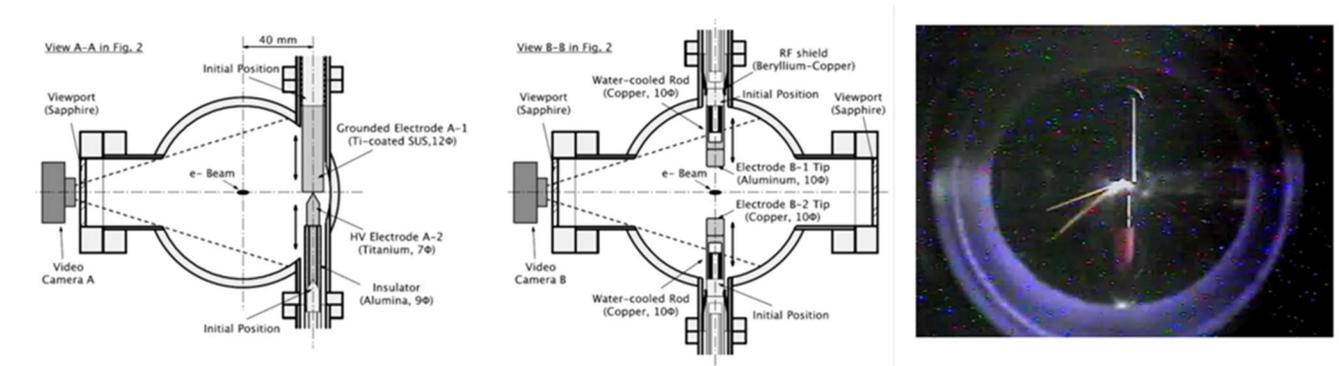

**Fig. 3**: Setup to demonstrate particulate capture by an electron beam (left and center). When high voltage is applied to the electrodes particulates can be generated. The capture of particulates was optically observed (right).

## 2.2 Beam lifetime limitations due to particulates

### 2.2.1 *HERA and DORIS*

In several accelerators sudden changes of the beam lifetime have been observed which could be traced to particulates moving in the vacuum chamber. Examples are HERA and DORIS at DESY and the Photon Factory at KEK. As shown in Fig. 2, sudden changes of the beam lifetime have been observed when the accelerators HERA and DORIS were operated with electrons. During operation with positrons this was not observed, see Ref. [5].

The events have been traced to positively charged particulates which have been emitted by the sputter-ion pumps integrated in the vacuum chamber. The negatively charged beam attracted the positive particulates. Eventually, beam and particulates collided leading to partial beam losses. Titanium particulates have been identified as the primary suspects. Even after the exchange of the chambers to a system with integrated NEG strips the effect could be observed albeit less pronounced. Consequently, HERA and DORIS have been operated with positrons most of the time after this discovery, see Ref. [6].

The effect was finally provoked in a dedicated experiment at the KEK Photon Factory. Electrodes were introduced in a special setup. When applying high voltage, particulates were generated. The subsequent capture of the particulate was visually observed as seen in Fig. 3. The details can be found in Ref. [7].

This example illustrates the need to evaluate whether the production of particulates during the accelerator operation can cause detrimental effects. Another obvious source of particulate production during accelerator operation are movable components like gate valves and beam diagnostics where the friction between metallic parts lead to particulate generation.

### 2.2.2 *LHC beam losses*

In 2010 and 2011 beam losses leading to 35 protection beam dumps were observed at LHC. With improved diagnostics information of about 7800 suspicious events was detected. The events were suspected to be caused by "Unidentified Falling Objects (UFOs).

A fraction of 6% of these events could be attributed to the injection kickers which are only 0.06% of the LHC length, see Ref. [8]. Other kickers were not showing is anomaly. Additionally, a clear correlation to the pulsing of the kickers could be established.

The explanation for the generation of the particulates is the vibration of the kicker due to the pulsed operation. This led to the release of aluminum oxide particles which were found in abundance after one item was removed from the accelerator. The other kickers in the machine have a metallic coating which prevents release of particulates.

### 2.2.3 *Optics components*

Particulate-related degradations can often be observed on optics components. In particle accelerators beam diagnostics sometimes requires sophisticated laser equipment. For highly brilliant electron beams photocathodes injectors are common which use high power lasers for illumination. In synchrotron radiation sources the transport of X-ray beams to the experiments requires many particulate-sensitive components. Mirrors and gratings can even be damaged as local heating at the particulates can lead to bending or melting.

# 3  Properties of particulates

Dust particulates originate from many sources. In the following some examples are given.
- Air pollution
    - Sahara sand
    - Diesel engines
    - etc.
- Fabrication
    - Machining
    - Drilling
    - etc.
- Assembly
    - Humans
    - Friction
- Operations
    - Friction
    - Charging
    - Aging

As can be seen, particulates contamination of vacuum component surfaces can occur at many stages from initial fabrication until the operation of the equipment in the accelerator.

To develop the proper countermeasure against contamination it is vital to understand properties of particulates. Even though the origins are diverse some properties are common.

## 3.1  Particulate size

Dust particulates have a wide range of sizes. Problems often occur with particulates of micrometer size or larger as described earlier in the case of the superconducting accelerating cavities. As indicated in Fig. 4 the size gives a rough estimate whether a particulate will be transported in air or other gaseous media. Whereas particulates smaller than about a micrometer are airborne, larger particulates will settle eventually on a surface due to the gravitational force. It should be mentioned here already, that even settling dust can remain airborne for a long time i.e. several hours.

**Table 1:** Specification of cleanroom classes, see Ref. [10]. The numbers is the accumulated count of particulates larger than the size given.

| Class | ≥0.1 μm | ≥0.2 μm | ≥0.3 μm | ≥0.5 μm | ≥1 μm | ≥5 μm | FED STD 209E equivalent |
|---|---|---|---|---|---|---|---|
| ISO 1 | 10 | | | | | | |
| ISO 2 | 100 | 24 | 10 | | | | |
| ISO 3 | 1,000 | 237 | 102 | 35 | | | Class 1 |
| ISO 4 | 10,000 | 2,370 | 1,020 | 352 | 83 | | Class 10 |
| ISO 5 | 100,000 | 23,700 | 10,200 | 3,520 | 832 | | Class 100 |
| ISO 6 | 1,000,000 | 237,000 | 102,000 | 35,200 | 8,320 | 293 | Class 1,000 |
| ISO 7 | | | | 352,000 | 83,200 | 2,930 | Class 10,000 |
| ISO 8 | | | | 3,520,000 | 832,000 | 29,300 | Class 100,000 |
| ISO 9 | | | | 35,200,000 | 8,320,000 | 293,000 | Room air |

The size of particulates can be used to filter them from gaseous and liquid media. Today, filters for virtually every process medium can be found. Typical examples are the filter-fan units (FFUs) which are used to filter air to make the setup of cleanrooms possible. For liquid media like water, alcohol or acids, depths or membrane filters are readily available. Nowadays, it is standard to filter particulates larger than 0.3 μm.

Cleanrooms are classified by the number of particulate of a certain size in a certain volume of air. The ISO 14644-1 classification is shown in table 1. For accelerator assemblies often class ISO 4 and ISO 5 are used. In mobile cleanroom tents in accelerator tunnels local environments of class ISO 5 can be established.

## 3.2 Particulate mass and transport

6.2 mg of dust particulates per m$^2$ per day are settling in german households due to the gravitational force. But the mass of each individual particle is very small. Thus small forces are sufficient to move them around. The molecules in gaseous media can move macroscopic particulates as is well-known from the experiment of the Brownian motion. When a medium like air is in motion, the particulates will move along with it, too. This is why in the earth's atmosphere Sahara dust can be transported several thousand kilometres by wind.

Even in vacuum system without turbulences time is needed for particles to settle on the surface. Figure 5 shows the number of particles which are transported when a vacuum system vented with ambient air (no particle filters used) is evacuated without turbulent flow after a waiting period of several hours.

Many particles can be measured, especially directly after the beginning of the pumpdown. These particles were still suspended in the gaseous media and had not settled. As can be seen from this measurement it can take several days for particles to stick to the surface. One should note that the Van der Waals forces help to keep particles sticking to the surface. The force needed to dislodge is thus somewhat larger than the gravitational force only.

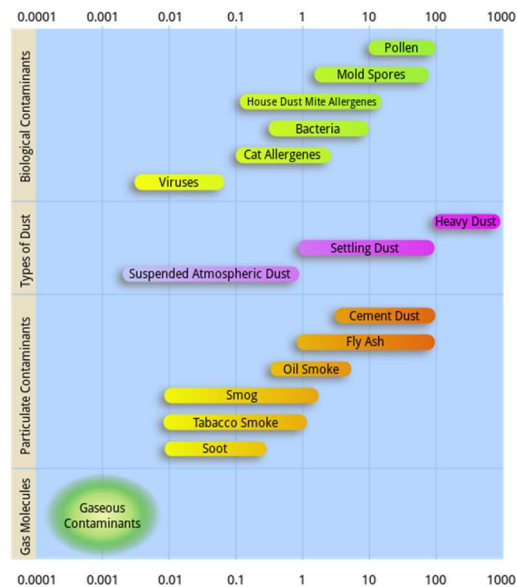

**Fig. 4**: Examples of particulates from various sources. The scale indicates the size in micrometers. Notably, humans as a source of contaminants like hair or flakes of the skin are missing. This diagram can be found in Ref.[9].

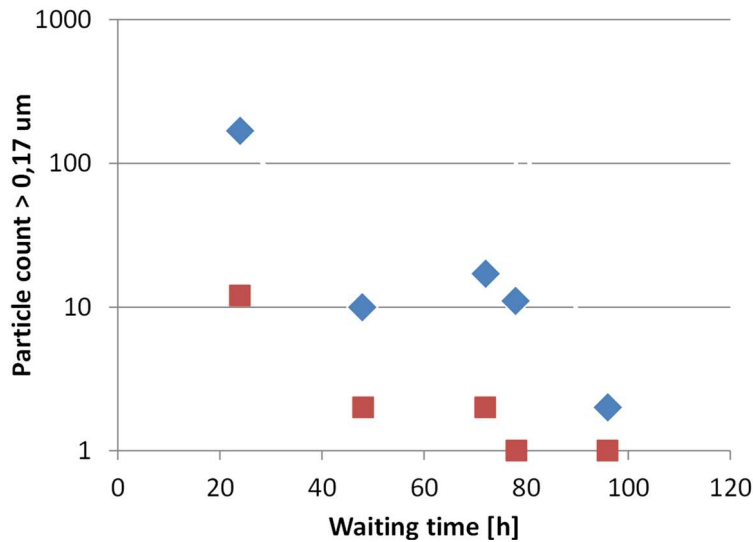

**Fig. 5:** Number of particles pumped after a vent of a vacuum system with an additional waiting time for the particles to settle. It takes several days for particles to stick to the surface so that they cannot be transported anymore. The measurement shows the number of particles when the pumpdown is started (blue diamonds) and about 10 minutes later (red squres) when the pressure in the systems was still about 10 mbar.

In contrast, particulates will simply fall down in UHV as every object with mass. Typical beampipes are traversed in a few ten milliseconds. Measurements have shown that up to about 1 mbar particles will not be transported and remain on the surface. Above this threshold, procedures in the installation and operation of "particle-free" vacuum systems have to be established to avoid particulate transport. A bad example is shown on Fig. 6. When an angle valve is opened at 10 mbar during a pump down of a vacuum system, particulates are dislodged from the surface due to vibrations and the turbulent flow and will be transported with gas in the system.

### 3.3 Particulate release from surfaces

Particulates can be dislodged from the surface if subjected to external forces. Mechanical vibrations can cause particulate release from a surface. An example is the operation of the angle valve which was mentioned before. Simple knocking with tools on metallic surfaces is sufficient to release the particles as well.

It is important to note that in vented vacuum systems vibrations can release particulates into the gas. Because of the Brownian motion, principally these particulates could be transported to every place on the inner surface. When preparing a "particle-free" vacuum system measures need to be taken to avoid excessive vibrations when the system is vented or pumped down at least to pressures of about 1 mbar.

Turbulent gas flow has a similar effect. Taking a book of a shelf where it was stored for some time leads to the reflex action of blowing away the dust layer on the top. Normal pump down and venting operations without flow restrictions will move particulates around in a similarly uncontrollable manner.

A third effect to efficiently release particles from a surface is to charge up the surface and the particles. This can be used for cleaning. For example, blowing ionized nitrogen on components is a standard cleaning technique used in cleanrooms. In contrast in particle accelerators, charging up due to the transported particles should be avoided. As outlined in section 2.2.1 charged particulates can cause detrimental effects.

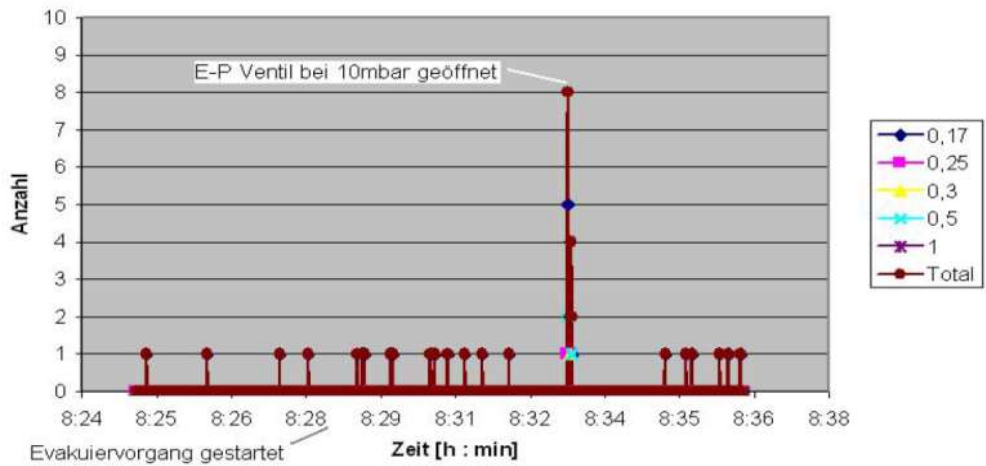

**Fig. 6**: During the normal opening procedures of angle valves particles can be dislodged due to vibrations or turbulent gas flow. In this case, the valve was opened a few minutes after the pumpdown procedure was started. The pressure in the system was about 10 mbar.

### 3.4 Particulate sources in accelerator operation

Accelerator operation inevitably leads to the production of particulates. Even when components have been thoroughly cleaned before and assembly procedures have been implemented to avoid particulate contamination it is important to understand that some particulates are being generated while an accelerator vacuum systems is operated. The operation means everything ranging from opening and closing of gate valves, moving diagnostics inside the vacuum system and even repairing or upgrading the accelerator complex.

#### *3.4.1 Moving components*

Moving components generate dozens to hundreds of particles even when they were cleaned before. Mainly this is due to the friction of the metallic parts that are part of movable devices. As an example gate valves can contain ball bearings which produce particles. Many diagnostics components like view screens have moving parts, too.

With some components their movement is less obvious. RF shielding is often implemented as sliding contacts which are compensating for example small length variations due to temperature changes. Again, particle generation is difficult to avoid. Wire scanners are specifically challenging as the thin wires, when hit by the particle beam can be destroyed with debris moving around uncontrollably.

Movements of components near critical areas like superconducting cavities should be minimised to avoid excessive particulate production. It is good practise to close valves in emergency e.g. an accidental venting needs to be contained or during maintenance with access of personnel, only. This allows a safe operation of the vacuum system while keeping the production of particulates at acceptable levels. In the FLASH accelerator at DESY this has been implemented successfully since many years.

#### *3.4.2 Pumps as sources of particulates*

The operation of a vacuum system includes the pump down and venting procedures. This can lead to particulate contamination. Some pumps produce particulates during operation. In mechanical pumps particle production is certainly possible like in scroll pumps. Scroll pumps are an attractive choice as a fore pump as they are "dry" pumps. Nonetheless, the tip seal wears during operation and the wear is nothing else than the generation of particulates. With regular maintenance and quality control, scroll pumps have not shown to be an issue and successfully in many projects.

In addition, the design of the mechanical roughing pump stations is crucial: An additional safety features like an automated valve for protection is mandatory in case the pump station fails. DESY has had good experiences during XFEL assembly and installation with a combination of scroll and turbomolecular pumps, which have been used in all roughing processes.

Sputter ion pumps have shown problems with particle generation when integrated into vacuum chambers e.g. HERA. Standard ion pumps have shown particle production especially during the initial start-up i.e. at comparatively high pressures. Measures exist to reduce particle transport e.g. optical shields or an optimised mounting position. Thus these pumps can be used for "particle-free" systems and they have been implemented in many accelerator systems.

The operation of NEG cartridges in "particle-free" systems is not yet that well known. Principally, tests on CapaciTorr pumps during several conditioning and activation cycles show that the number of particulates generated is at acceptable levels after 4 repetitions as shown in Ref. [11]. A pre-conditioning on test stands seems feasible, so that during operation particulates will not be produced. In addition, newer materials might further address the issue of particles production.

## 4  Cleaning and keeping components clean

Particulates can be removed from surfaces as mentioned before. Methods for surface cleaning rely on several the effects mentioned: Vibrations, turbulent flow of liquid or gaseous media and charging. In some cases additional cleaning can be achieved by the adding detergents which increase the chemical solubility of components. This improves the cleaning efficiency for film-like contaminations like grease which are often not particulates in the stricter sense.

### 4.1  Ultrasonic cleaning

Several cleaning methods are efficient procedures to clean components for both vacuum and "particle-free" specifications. A very well-established tool is ultrasonic cleaning. Components are immersed in a bath of water with added detergent. High frequency pressure waves are generated by ultrasonic transducers. This leads to cavitation in the solution on the surface of the component. The cavitation exerts sufficient force to remove particles from the surface. Typically, the liquid solution is also agitated to transport particulates released away from the surface.

Ultrasonic cleaning is followed by ultra-pure water rinses to remove residues of the detergent. In a final step components are often stored in special drying areas or cabinets to quickly remove the water film to avoid drying stains and oxidation.

These procedures can produce very good vacuum properties with a low contamination of hydrocarbons on the surface as has been shown for FLASH and the European XFEL. Components subjected to this cleaning show a very low particle count when crosschecked by ionized nitrogen blowing.

### 4.2  Ionized nitrogen blowing

Blowing components with ionized nitrogen is a standard cleanroom technique for removing dust particulates. A nitrogen gas stream is ionized in an ionizing "gun" by an electric discharge. When directed to contaminated surfaces particulates and the surface will charge up and a force will be exerted on the particulate and release it. The gas stream will then transport the particulates away.

On components pre-cleaned with ultrasonic cleaning this blowing can serve as a quality check. The particles blown away can be (partially) collected with a particle counter which has to be located at a reasonable position i.e. the downstream gas stream. When no particles are counted the component can be considered clean. When this process takes excessively long i.e. a few minutes, the cleaned component should go back to ultrasonic cleaning as this is usually more efficient.

Some sensitive components might not be cleaned by ultra-sound cleaning. In that case ionized nitrogen blowing is the only tool. It should be noted that this operation can be very time-consuming and tedious.

### 4.3 High pressure water rinsing

A technique specifically built for superconducting niobium cavities is the high pressure, ultra-pure water rinsing. After the initial ultrasonic cleaning, (low-pressure) ultra-pure water rinse and drying in controlled particle-free atmosphere, normally cavities have to be equipped with additional components like antennas. To reduce the risk of particles introduces during these assembly steps the cavity is rinsed again with a jet of high pressure pure water. With this procedure the mechanical force of a water jet to remove the particles.

It is preferable to ultrasonic cleaning at this stage, as the there is a higher risk of having detergent residues in parts of the cavity which are no accessible any more. For niobium, water does not immediately lead to strong oxidations, so that a final water rinse produces a good superconductor surface.

### 4.4 Dry-ice cleaning for normalconducting RF structures

RF structures based on copper are much more sensitive to oxidation when prepared with wet cleaning techniques. Often discolorations due to oxidation can be observed. Dry-Ice cleaning for normalconducting RF structures is a good alternative, as described in Refs. [12, 13]. In this case a jet of carbon dioxide snow is directed to the surface, see Fig. 7. The jet is stabilized by an external gas flow of nitrogen. When the jet hits the surface, rapid cooling can embrittle contaminations, high pressure of the jet exerts shearing forces and the expanding volume due to sublimation rinses away particulates. Additionally, the liquid phase of the carbon dioxide can help to dissolve non-polar film contaminations like hydrocarbons.

As the surface remains dry i.e. without a water film, drying residues and oxidation can be avoided. Some attention is required to keep the component under cleaning warm enough to avoid condensation due to the humidity of the surrounding air.

This method has been successfully applied to both superconducting and normalconducting structures. Due to its "dryness" the application to components based on copper is very attractive. RF photoinjector guns are being cleaned systematically at DESY.

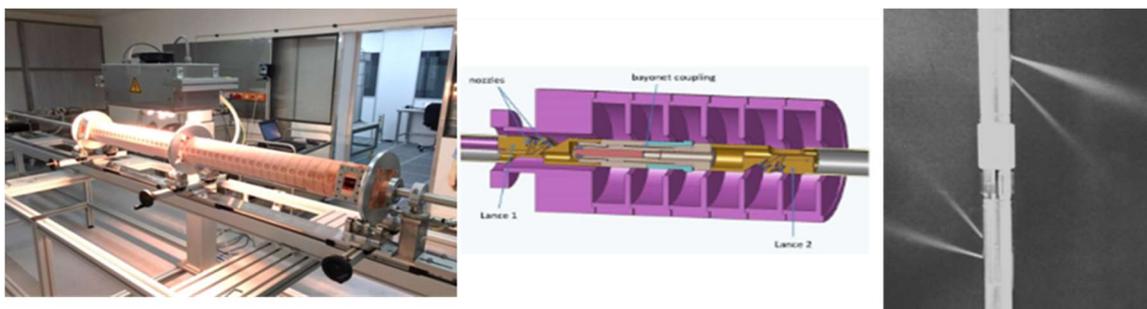

**Fig. 7:** View on the setup for cleaning accelerating structures with dry-ice snow. (left). An lance is inserted which allos to clean the structure horizontally (center) .View of the nozzle during operation (right).

### 4.5 Storage and transport of cleaned vacuum components

"Particle-free" vacuum components need to be transported to the area were they are going to be installed to an accelerator. After cleaning for particles, the transport can lead to a recontamination. During transport vibrations are sometimes difficult to avoid.

Ideally, the components should be transported under vacuum inside. As outlined above particles will stick to the surface and not move around. If they are dislodged due to vibrations, usually they will fall down again. As there is no gaseous media, they will not be transported around in the vacuum volume. The clear downside of this approach is that much more effort is needed in terms of equipment like additional valves and gauges for monitoring. Typically, an additional venting and pumpdown is needed just before installation into the accelerator complex. For critical components like accelerator modules or cavities this additional effort is well justified.

For simpler components, a simple backfill with particle filtered nitrogen is a reasonable solution. Simple bellows or gaskets can be quickly re-cleaned by ionized nitrogen blowing and cross-checked before the installation. Nonetheless, the risk of re-contamination of areas that have been cleaned is significantly larger, as the gas will facilitate the transport of particulates.

To avoid a time-consuming re-cleaning of the components in the installation area it can be useful to (double-)bag the outside in antistatic foil. As the components often are pre-assembled in cleanrooms, the outside has already been cleaned reasonably. Some cleaning is inevitable to reduce the risk of cross contamination between outside and inside during final assembly e.g. mounting connecting bellows.

### 4.6 Avoiding particulate transport during pumpdown and vent

Avoiding turbulence is essential to not move particles. Any transport of particulates during venting or pump down can lead to a relocation to a part of the vacuum were a particulate can do damage. Therefore, laminar gas flow is an important tool to avoid a surface contamination during both venting and pump down of the "particle-free" vacuum system. It is clear however, that if particles are released into the laminar flow they will of course be transported. Avoiding turbulences and avoiding vibrations have to be implemented alongside to make sure that the gas stream needed for venting or during pump- down will not lead to a contamination.

Measurements as described in Ref. [15,16] with an in-vacuum particle counter have shown that a particle transport can be avoided when either

the gas flow ≤ 3 $l_n$/min, or

the pressure < 1 mbar.

Automatic pumping and venting units have been developed at DESY to ensure a constant flow of 3 $l_n$/min of nitrogen by means of mass flow controllers. These units have been widely used for XFEL for the assembly of the cavities to accelerator modules as well as for the installation of the modules into the accelerator.

## 5 Design considerations and accelerator layout to avoid particulates

If a "particle-free" vacuum system is needed for an application, the design and the layout has to reflect this and thus needs to be taken into account during every stage of the project. There are three topics which will be briefly addressed here: Mechanical design of individual components, Segmentation to facilitate installation and operation, and finally Tunnel layout and local cleanroom environments.

## 5.1 Mechanical design and particulates

Particulate contamination can be reduced by a proper mechanical design and the optimal choice of off-the-shelve components. The main point is to ensure that efficient cleaning can be done. As wet cleaning in an ultrasonic bath is extremely efficient components should be designed in a way that liquids can be easily drained during the different stages of rinsing. Avoiding dead spots is important as these areas increase the risk to be insufficiently cleaned both for particulate and chemical residues e.g. drying stains. Ideally, a continuous uni-directional flow can be established so that particulates released from the inner surface can be washed away.

Some cleaning procedures require the insertion of a lance for optimal results e.g. dry-ice cleaning. An adaption to the required sizes for penetrations can significantly reduce processing times or avoid expensive adaptions of the cleaning systems.

As outlined before, the minimum requirement is to allow for a gas stream ionized nitrogen to access the surface. Blowing components is a tedious business. Therefore, an optimized design has to take into account this and should make sure that the processing times are acceptable. In some cases using dummy pieces is quite useful to see where particulates would be trapped potentially.

An obvious example for an off-the-shelve component very difficult to be cleaned is a welded bellows. It is much preferable to use hydroformed bellows wherever possible due to the easier cleaning process that can be applied. For fabrication techniques, full penetration welds are preferable. Gaps or fissures are difficult to clean for particulates and liquid process media are difficult to rinse away.

Assembly processes in cleanroom environments take advantage of a directed particle-filtered air stream in the cleanroom. Typically, the stream is laminar directed from the ceiling to the floor. The mechanical design should allow for having an optimal assembly where the operator with his tooling can access the area of work from the downstream side. The clean flow should have access to the area of work with no obstruction.

It is a clear recommendation to discuss the mechanical design of components to be installed in a "particle-free" area with the responsible parties for cleaning at a very early stage.

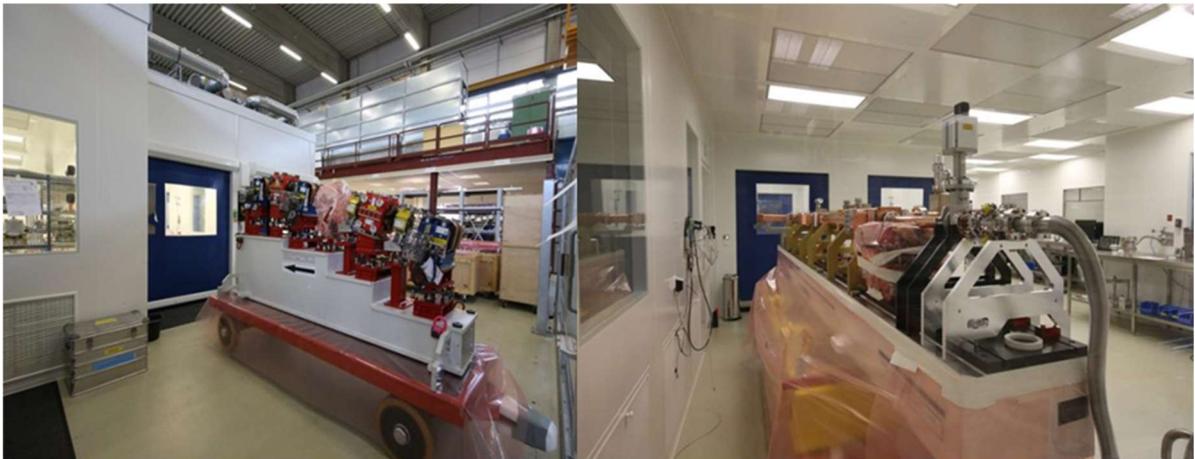

**Fig. 8:** Outside view of the DESY machine vacuum group cleaning facility. A girder is awaiting transport to the accelerator tunnel (left). View of the girder assembly area in the cleanroom. This part has been added to the facility described in Ref [14].

## 5.2 Segmentation

Installation of a "particle-free" vacuum system requires setting up controlled, clean areas. While the requirements for this installation are being discussed in the next subsection, it is sufficient to say here that it is a significant effort in an accelerator tunnel environment where compromises are often unavoidable.

Every single flange connection needs to be done under cleanroom conditions. Moving around a cleanroom tent in an accelerator tunnel is not just a lot of time-consuming work, but often interferes with the space needed for other installation like cabling etc. .

Thus for the assembly of "particle-free" sections it is much preferable to have as many components pre-installed and connected as subsections in a well-controlled dedicated cleanroom facility like the one shown in Fig. 8. The number of connections which need to be made in the accelerator itself can be drastically reduced by setting up girders which are subsequently transported into the tunnel. For the European XFEL a girder length of about 5 m length was chosen. About an average of 4 to 5 components were assembled on a girder in an above ground cleanroom facility. As only the connection between two of these girders needed to be made in the tunnel by insertion of a comparatively simple bellow, the number of connections in the tunnel was reduced by a factor of 3-4.

Using the same bellow connection over and over again allowed for the design of a dedicated small cleanroom environment with a smaller footprint than a large multi-purpose cleanroom tent. Other trades were less affected by this installation method. In addition, a reliable, efficient cleanroom setup improves the overall quality of the installation work as it needs less time and allows the installation team to focus on the critical assembly steps.

For the European XFEL, more than 1.5 km of vacuum components have been cleaned in the infrastructure of the DESY machine vacuum group. A large variety of objects ranging of simple tubes for beam transport and large aperture bunch compressor chicane chambers to complex diagnostics components have been handled. The DESY infrastructure described in Ref. [14] has been extended for additional assembly space allowing the handling of the 5m girders of the European XFEL accelerator.

The second need for segmentation is the more general requirement of a vacuum system to ensure that accidents can be contained, leak searches span reasonable length and pump-times are acceptable. This is realized with placing gate valves and fast shutter at the appropriate positions.

## 5.3 Tunnel layout and mobile cleanrooms

The setup of cleanroom environments can be challenging and time-consuming. Thus it is mandatory to think about the installation process and the associated equipment early in the accelerator design. The "particle-free" installation area requires space for several reasons:

- It is obvious that the FFU, power supply etc. need space
- Establishing a laminar flow for a certain area requires avoiding obstructions in a certain volume between the FFU and the work area.
- Work area for the operators required to perform the task
- Space is needed for operators to dress according to what is needed
- An area for storage of clean tooling inside the cleanroom environment.
- Sufficient lighting
- Additional tools or fixtures to simplify the mounting require space
- Space for the vacuum operations: pump station, leak search, nitrogen dewars etc.

These spaces should be considered during tunnel layout to make sure getting optimal conditions for particle-free assembly.

Several solutions exist. A simple FFU with some plastic foil as a curtain can give already good results. Often, this is the most flexible solution. It is very useful when the work area is small.

When assemblies have to be repeated multiple times it is advantageous to adapt a cleanroom to the assembly procedure. This will reduce the time to setup an environment significantly and results in more repeatable and thus reliable handling. An example is given in Fig. 9.

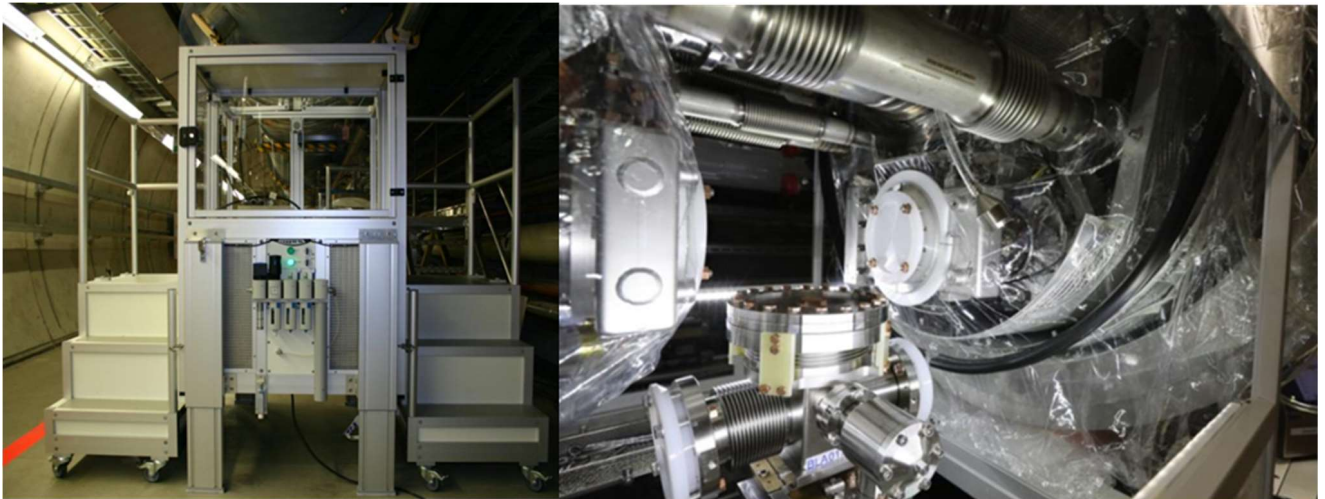

**Fig. 9:** Side view of the specialised cleanroom for absorber assembly of the European XFEL (left). View inside the work area (right). The absorber is still in the parking position below the beam line. Plastic caps for protection during transport are still mounted. A part of the cryogenic pipes can be seen above.

In the case of the European XFEL a broad band higher order mode absorber interconnects two accelerator modules. The absorber weighs about 10 kg. About hundred of these assemblies had to be done.

The adapted cleanroom features a laminar flow from bottom to top. This was chosen as the cryogenic piping had to be welded before the absorber assembly. In a top-down configuration the laminar flow would have been disturbed significantly by several tubes. The clean volume is enclosed by a rigid housing and a disposable foil. The rigid part reduces the setup time for the cleanroom. The disposable foil is placed on the accelerator module cryostat side before welding as a barrier to the "dirty" interior of the cryostat and is removed after the absorber assembly.

A movable support can position the absorber to the correct space between two gate valves, which simplifies the assembly significantly. A particle counter for quality control and ionizing nitrogen gun are located inside the clean area.

Operators can access the work area from two sides. They have only the lower part of the arm in the clean area and do not need to dress completely. Finally, the stairs to each side can be removed easily during waiting times like pumpdown to allow for transports.

The adapted cleanroom has worked very well and allowed for very repeatable working conditions. So far, there are no indications, of deteriorations of accelerator performance due to assembly procedures.

# 6 Basic rules for a "particle-free" accelerator

The above can be summarized in a few rules:

1. Avoid particulates at every stage
   - This needs to be considered in the early phase of the mechanical design
   - Ensure setup of proper cleanroom environments
2. Remove particulates at every stage possible
   - Reliable cleaning procedures need to be established
3. Do not produce particulates especially during installation and operation
   - Proper choice of materials
   - Reduce the number of critical components (e.g. movable screens) in particle-free areas
   - Pre-configured sub-assemblies help to reduce the number of assembly steps in a difficult environment
4. Never transport particulates
   - A controlled gas flow is mandatory for pumpdown and venting operations
   - Transport under vacuum conditions might be justified for complex components

The above rules describe an ideal situation. It is obvious that conflicting interests need to be considered. In most cases reasonable compromises can be achieved.

# 7 Summary

Particulates can cause severe problems in accelerator operation like performance degradation or availability issues. Sources for particulates are diverse. Several possibilities exist to implement environments with a low particulate contamination. Filters for air and liquid media are readily available nowadays. Contamination control has to start with a proper mechanical design and accelerator layout. Particle transport due to venting and pump down can be avoided by using flow restrictions. Large-scale accelerator infrastructures like the European XFEL can be built "particle-free" and show no sign of performance degradation up to now.

# Acknowledgement

The author would like to thank the dedicated staff of the DESY machine vacuum group for their outstanding work in developing the technology of UHV system with very low particulate contamination.

# References

[1] https://www.wdr.de/tv/applications/fernsehen/wissen/quarks/pdf/Q_Staub.pdf, last accessed 17 April 2018

[2] H. Padamsee, The science and technology of superconducting cavities for accelerators, Supercond. Sci. Technol., 14 (2001), R28 –R51

[3] The European X-Ray Free Electron Laser Technical Design Report, http://xfel.desy.de/technical_information/tdr/tdr/ , last accessed 17 April 2018

[4] P. Schmüser et al., Superconducting TESLA cavities, Phys Rev ST-AB, VOLUME 3, 092001 (2000)

[5] Daren Kelly, Many-Event Lifetime Disruption in HERA and DORIS , DESY HERA 95-02

[6] Alexander Kling, Dust particles in HERA und DORIS, Proceedings of EPAC 2006, Edinburgh, Scotland, TUPLS002

[7] Yasunori Tanimoto, Experimental demonstration and visual observation of dust trapping in an electron storage ring, PRST-AB 12, 110707 (2009)

[8] B. Goddard et al., Transient Beam Losses in the LHC Injection Kickers from Micron Scale Dust Particles, IPAC2012, TUPPR092

[9] https://en.wikipedia.org/wiki/Particulates, last accessed 17 April 2018

[10] https://en.wikipedia.org/wiki/Cleanroom , last accessed 17 April 2018

[11] S. Lederer et al., Particle Generation of CapaciTorr Pumps , Proceedings of IPAC2017, Copenhagen, Denmark, WEPVA048

[12] A. Brinkmann, J. Iversen, D. Reschke, J. Ziegler, "Dry-Ice Cleaning on SRF-Cavities", in Proceedings of EPAC 2006, Edinburgh, Scotland, June 2006, paper MOPCH154, pp. 418 – 420.

[13] Brinkmann, J. Ziegler, Dry-Ice Cleaning of RF-Structures at DESY Proceedings of LINAC2016, East Lansing, MI, USA, MOPRC032

[14] K. Zapfe, U. Hahn, M. Hesse, H. Remde, A Cleaning Facility to Prepare Particle-Free UHV-Components, Proceedings of the 11th Workshop on RF Superconductivity, Lübeck/Travemünder, Germany, TUP50, pp. 457 - 460

[15] K. Zapfe, J. Wojtkiewicz, Particle Free Pump Down and Venting of UHV Vacuum Systems , Proceedings of SRF2007, Peking Univ., Beijing, China, WEP74

[16] M. Böhnert, D. Hoppe, L. Lilje, H. Remde, J. Wojtkiewicz, K. Zapfe (2009) Particle Free Pump Down and Venting of UHV Vacuum Systems. Proc. of the 14th Workshop on RF Superconductivity, Berlin, 2009, THPPO104